\documentclass[11pt]{article}
\usepackage[margin=2.5cm]{geometry}
\usepackage{color}
\usepackage{xcolor}
\usepackage{graphicx}
\usepackage{amsmath}
\usepackage{amsfonts}
\usepackage{soul}
\usepackage[normalem]{ulem}
\usepackage{lipsum}
\usepackage{footnote}
\usepackage{bibentry}
\usepackage{float}
\usepackage{authblk}
\usepackage{booktabs}
\usepackage[hidelinks]{hyperref}
\usepackage{silence}
\usepackage{cite}

\begin{document}

 \title{Rapid non-destructive inspection of sub-surface defects in 3D printed alumina through 30 layers  with 7 $\mu$m depth resolution}

\author[1,*]{C. Lapre}
\author[2]{D. Brouczek}
\author[2]{M. Schwentenwein}
\author[3]{K. Neumann}
\author[3]{N. Benson}
\author[1,4]{C. R. Petersen}
\author[1,4,5]{O. Bang}
\author[1,4]{N. M. Israelsen}
\affil[1]{DTU Electro, Department of Electrical and Photonics Engineering, Technical University of Denmark, Ørsteds Plads 343, 2800 Kgs. Lyngby, Denmark}
\affil[2]{Lithoz GmbH, Mollardgasse 85a/2/64-69, 1060 Vienna, Austria}
\affil[3]{airCode, Bismarckstr. 142, 47057 Duisburg, Germany}
\affil[4]{NORBLIS ApS, Virumgade 35D, 2830 Virum, Denmark}
\affil[5]{NKT Photonics A/S, Blokken 84, 3460 Birkerød, Denmark}
\affil[*]{Corresponding author: C. Lapre, corla@dtu.dk}
\date{}

\maketitle

\section*{Abstract}
The use of additive manufacturing (AM) processes for industrial fabrication has grown rapidly over the last ten years. The most well-known AM technologies are fused deposition modelling and stereolithography techniques. One particular industry where 3D printing is advantageous over traditional fabrication techniques is within ceramic components due to its flexibility. To establish a new and improved level of print quality and reduce resource consumption in the 3D printing ceramics industry, there is a need for fast integrated, sub-surface and non-destructive inspection (NDI) with high resolution. Several techniques have already been developed for high-resolution NDI, such as X-ray computed tomography (XCT), but none of them are both fast, integrable, and non-destructive while allowing deep penetration with high resolution. \newline
In this study, we demonstrate sub-surface monitoring of 3D printed alumina parts to a depth of $\sim$0.7 mm in images of 400$\times$2048 pixels with a lateral resolution of 30~$\mu$m  and depth (or axial) resolution of 7$~\mu$m . The results were achieved using mid-infrared optical coherence tomography (MIR OCT) based on a MIR supercontinuum laser with a 4$~\mu$m center wavelength. We find that it is possible to detect individual printed ceramic layers and track predefined defects through all four processing steps: green, preconditioned, debinded, and sintered.  Our results also demonstrate how a defect in the green phase could affect the final product. Based on the understanding of how defects develop in maturing printed parts, we pave the way for NDI integration in AM, which can be combined with artificial intelligence and machine learning algorithms for automatic defect classification in volume production of a new standard of high quality ceramic components.

 \section{Introduction}

The additive manufacturing (AM) industry has grown tremendously over the last ten years, in part because it allows to fabricate components that are impossible to make with more traditional techniques like injection molding. There are different technologies of AM \cite{Wang2017,Shahrubudin2019}. The most common is  fused deposition modelling (FDM), where the printer deposits the filament material layer by layer, for example using plastics  or concrete \cite{Benamira2023,Xiao2021}. The second most common technology is stereolithography (SLA), which uses a liquid ultraviolet-curable photopolymer resin and a ultraviolet (UV) light source to build parts one layer at a time \cite{Quan2020}. This technology has a better resolution than FDM, $\sim$50$~\mu$m on average compared with the best achievable resolution of FDM of $\sim$100$~\mu$m. In the area of ceramics fabrication, the major fabrication technology is AM using SLA 3D printing techniques \cite{Su2008,Hassanin2021}. During the AM process, it happens that defects are created due to an inconsistent material deposition or due to an external contamination, resulting in the presence of air pockets. While such defects  affect part performance, they remain invisible to surface inspection. To understand, and reduce volume defects, there is a need for fast and high resolution sub-surface non-destructive and non-contact imaging, which can be integrated into the printer to reach higher performance-defined print specifications. 
\newline

Another motivation is directly linked to reducing waste and costs during manufacturing. An industry ready sub-surface scanner can provide novel in-house understanding of defect formation, thereby reducing the number of defects, while revealing low-quality prints in the first process step, potentially saving a vast amount of energy in further thermal post-processing.\newline

There are many different ways to perform in-depth NDI of ceramics, the most well-known being ultrasound imaging, infrared thermography and  X-ray computed tomography (XCT). Another emerging technology is terahertz imaging, which is promising in terms of material penetration but unsuitable for 3D imaging due to low speed and low spatial resolution. The limitation of ultrasound is foremost the requirement of a contact medium (e.g. gel or water) and the constraint on spatial resolution, making it unable to see microscopic details. For XCT imaging, the requirement of rotating either detector or sample is the main barrier in industry and its use of hazardous illumination further complicates operation. Infrared thermography is becoming more applied in industry, being fast and suitable for large area scans. The in-depth information provided is only a low resolution en-face average projection of a volume and the sample has to be able to withstand a significant amount of heat. Access to cross-sectional images is absent, while detail is typically on the millimeter scale. None of these NDI techniques are easy to integrate into the manufacturing  process in combination with providing large scan areas and microscopic volume mapping with high accuracy and sensitivity \cite{Steckenrider2013,Duan2019,Zhao2021}. 
\newline

Optical coherence tomography (OCT) is a laser based technology that is fast, non-contact, non-destructive, and allows for imaging of very complex structures \cite{Drexler2001,Israelsen2018}. The technology has demonstrated its efficiency for more than 30 years \cite{Fercher1988} in ophthalmological clinics and in biology, e.g., for delineating cellular interfaces and imaging the human retina and cancerous cells \cite{Fujimoto2000,Boehringer2009,Walther2011}. In contrast to ultrasound, OCT allows only for reaching a few millimetres in depth, however, with a much better resolution of 1-15~$\mu$m.  
\newline

Although OCT is well established in the biomedical field, the technology is still relatively new to the industrial domain.  Most OCT techniques use near-infrared (NIR) wavelength laser sources with center wavelengths around 600-1300~nm and at these wavelengths inorganic materials, such as ceramics, are opaque due to strong scattering. With a laser source above  3~$\mu$m, light scattering is strongly reduced, which allows for much deeper penetration. For this reason, mid-infrared (MIR) OCT has been developed recently, unravelling new NDI applications. Examples include sub-surface monitoring of marine coatings \cite{Petersen2021}, paper quality inspection \cite{Hansen2022}, NDI of wind turbine blade coatings  \cite{Petersen2023}, credit cards, and also initial  research on ceramic characterisation \cite{Zorin2022}. \newline 

In this paper, we demonstrate MIR OCT as an NDI scanner tracking sub-surface microscopic defects throughout all processing steps of 3D printed alumina parts. Predefined defects in AM alumina are imaged in four stages of the post-printing for three different print configurations. This project is a part of the  Horizon Europe project ZDZW \cite{ZDZW_project} targeting zero defect, zero waste manufacturing in Industry 4.0.

\section{Material and method}

\subsection{The alumina samples}\label{alumina_section}

\begin{figure}[H]
\centering\includegraphics[width=15cm]{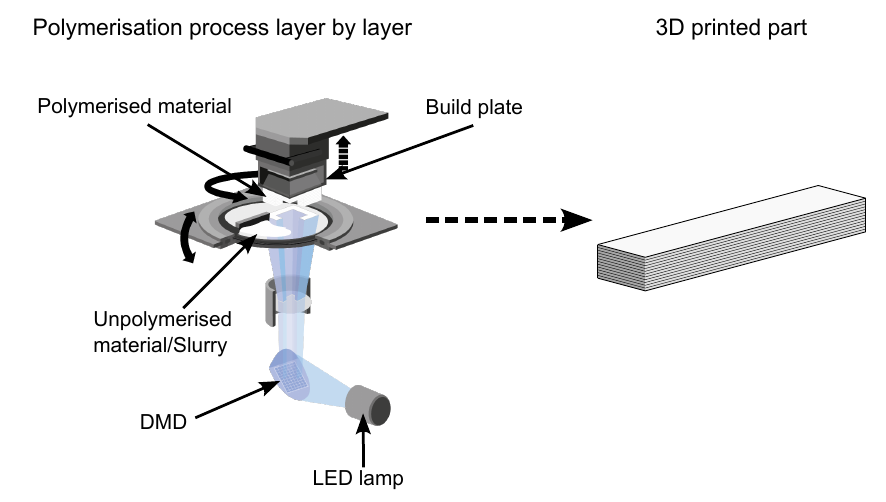}
\caption{Schematic of the 3D printing procedure, based on a commercial LITHOZ 3D printing machine, DMD : digital mirror device. A 3D file is created and designed with a CAD software and then sent to the 3D printing machine. Inside the printer, a build plate is dipped into a rotatable vat where the photosensitive ceramic suspension (slurry) is distributed for each new layer. Through an optical system, the material is selectively exposed via the LED lamp and cured by photopolymerisation at these specific exposure areas, creating a volumetric part layer by layer. The final product is a part designed and printed in 3D.}
\label{print}
\end{figure}

 The AM approach was based on a commercial LITHOZ 3D printing machine using lithography-based ceramic manufacturing technology (LCM), with applications in for example the aerospace and medical industries. The fabrication process  resembles the SLA process, as depicted in Fig.~\ref{print}. Essentially, the quality of the final product will depend on the resolution of the digital mirror device (DMD) after the LED lamp and on the process parameters, the print method and the material properties of the slurry used for the printing \cite{Conti2020,LithozGmbH2022}. \newline

After printing, the fabrication is not complete and requires subsequent cleaning to remove residual uncured material from the surface of the part. Following cleaning, the part needs to undergo thermal post-treatment to remove the remaining organic material, and then sintering is performed to finally densify the ceramic product. This process reduces the size of the sample by around $\sim$20 \%.\newline

The different procedures of cleaning and thermal post-processing for the 3D printing of ceramics are conventionally called the following \cite{Ferkel1999}: 
\begin{enumerate}
    \item Green bodies: component directly after printing and cleaning.
    \item Preconditioned components: after a thermal treatment at 120°C.
    \item Debinded components: after a thermal treatment at 600°C.
    \item Sintered components: after a thermal treatment at 1650°C.
\end{enumerate}

The preconditioned and debinded phases remove all the remaining organic material still present inside the part. The sintered phase will densify the parts, causing the layer structure to disappear.\newline

\begin{figure}[H]
\centering\includegraphics[width=15cm]{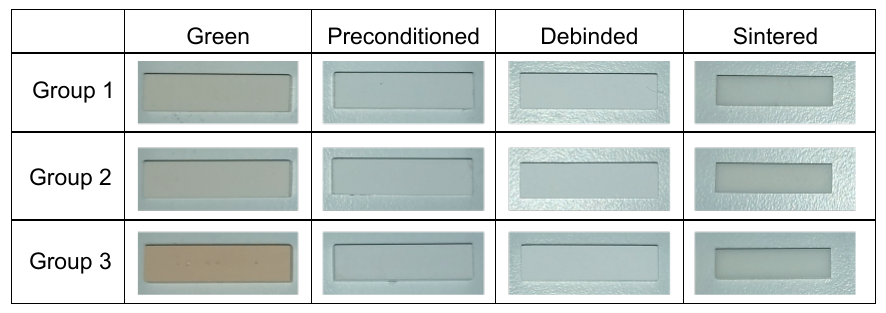}
\caption{Photographs of the investigated sample prints ordered by print recipe (Group 1-3) and production phase, sample size after printing (green) $4.96 \times 19.8 \times 2.56\,\mathrm{mm^3}$, dimension of sintered sample (sintered) $4.02\times 16 \times 2.01\,\mathrm{mm^3}$.}
\label{sample}
\end{figure}

We characterized three different groups of printed alumina ceramic samples representing three print configurations. The intention of these samples was to monitor the impact of relevant defects on the internal structure of ceramic antenna modules during the printing process. The layer thickness (green phase) is 25~$\mu$m and the total number of layer inside each sample is 102. The defects are cube air pockets with designed dimensions of $800 \times 800 \times 800\,\mathrm{\mu m^3}$ (green phase) and final dimensions of $648.8 \times 648.8 \times 628.4\,\mathrm{\mu m^3}$ (sintered phase). As the dimensions of the preconditioned and debinded samples were not characterised, we cannot provide them. The difference between these three groups is the material used, and the positioning of the intentional defects. Table~\ref{print_method} presents the different print methods of each sample: the position of defects during printing, and the slurry material used. Each group of samples contained one example of each processing stage, being green, preconditioned, debinded and sintered, as listed above and presented in Fig.~\ref{sample}.\newline

\begin{table}[H]
    \centering
    \begin{tabular}{|c|c|c|}
   \hline
        Group 1  & Group 2   & Group 3  \\ %(A23043) (A23040) (A23047)
               \hline
         beige slurry & beige slurry  & red slurry \\
         \hline
       \includegraphics[width=0.3\textwidth]{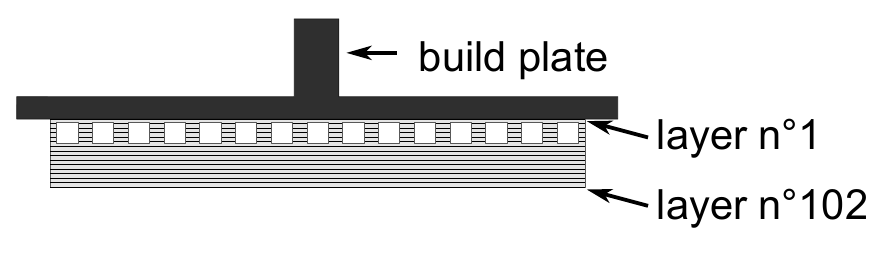}  &   \includegraphics[width=0.3\textwidth]{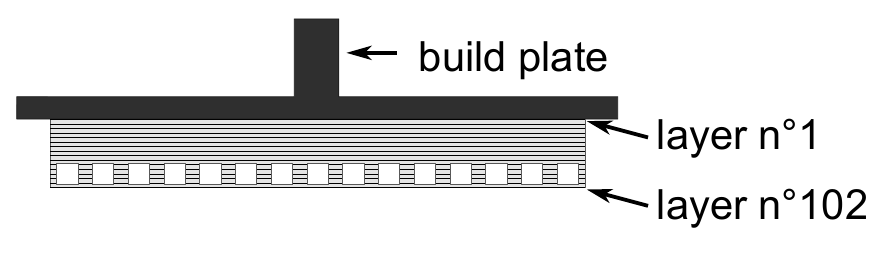} &  \includegraphics[width=0.3\textwidth]{schema_print_b.pdf}\\
        \hline
       \includegraphics[width=0.3\textwidth]{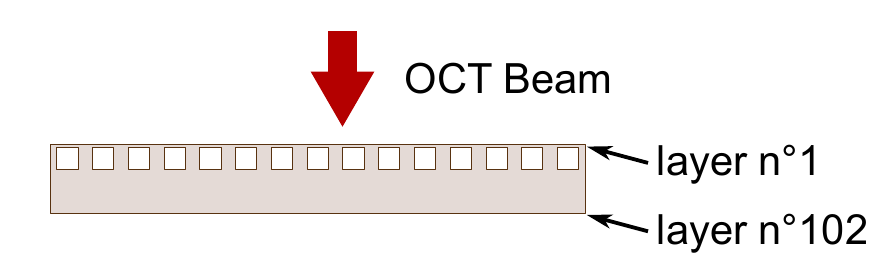}  &   \includegraphics[width=0.3\textwidth]{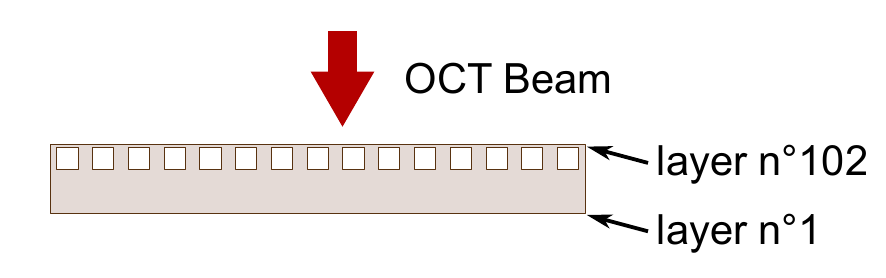} &  \includegraphics[width=0.3\textwidth]{schema_mesure_b.pdf}\\

         \hline
    \end{tabular}
    \caption{Summary of printing methods used for the different sample groups and the direction of the OCT beam. The white boxes represent the positions of the intentional defects inside the samples. For imaging Group 2 and 3, samples are rotated 180$^\circ$ relative to the printing direction to minimise the path length to the defects.}
    \label{print_method}
\end{table}

\subsection{MIR OCT} \label{mir_oct_section}

\begin{figure}[H]
\centering\includegraphics[width=15cm]{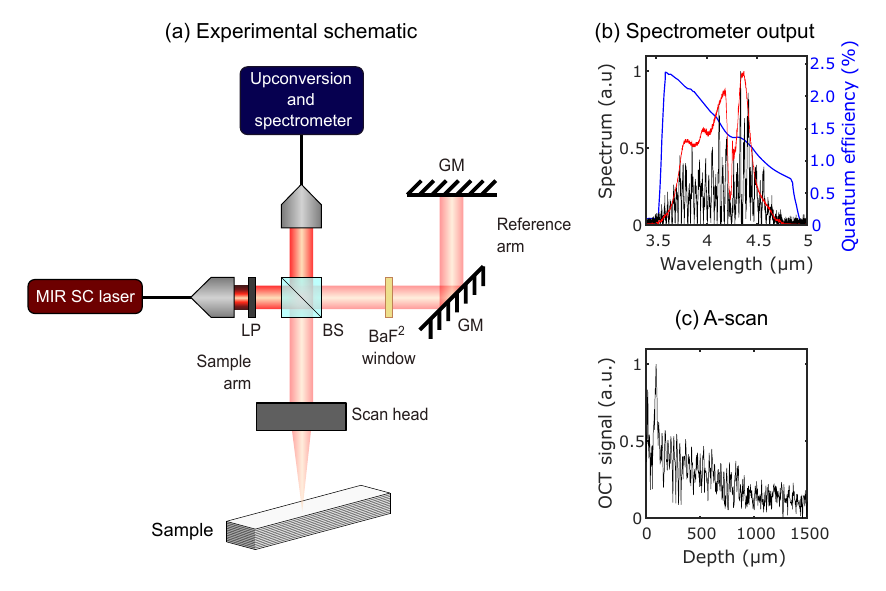}
\caption{(a) Schematic of the MIR OCT system, LP: 3.2~$\mu$m long pass filter, BS: beam splitter, SC: supercontinuum, GM: gold mirror. (b) Reference spectrum in red, measured interference spectrum in black, upconversion efficiency in blue, and (c) the corresponding calculated A-scan.} 
\label{oct}
\end{figure}

The NDI scanner is depicted in Fig.~\ref{oct}. The MIR OCT system was based on an in-house fabricated supercontinuum (SC) fibre laser  covering a continuous spectrum from 1 to 4.6 $\mu$m \cite{woyessa2021power}. The laser source was coupled into a free-space Michelson interferometer. Just after the input of the interferometer, the light was filtered with a long-pass filter to remove the light below 3.2 $\mu$m. The interferometer splited the light in two with a beamsplitter, and then injected the light into two different arms. One arm was the sample arm, and the other was the reference arm. \newline

In the sample arm, scanning was achieved using a two-axis silver-coated mirror galvanometric scanner coupled to a barium fluoride (BaF$^2$) plano-convex lens with a focal length of 30~mm. The average power inside the sample arm was 18~mW. The reference arm had two gold-coated plane mirrors to align or mis-align the reference beam. A slight misalignment allowed to attenuate the power of the reference beam. As the lens inside the sample arm introduced chromatic dispersion, a BaF$^2$ window inside the reference beam path was installed to compensate the dispersion. The rest of the dispersion was compensated numerically \cite{Wojtkowski2004}.\newline 

After backreflection, the two laser signals were then combined by the beam splitter to interfere. The interference signal was converted from 4 $\mu$m to 0.8 $\mu$m via an upconversion system based on sum frequency generation \cite{Rottwitt2014}, to benefit from the sensitivity, resolution, and speed of a NIR spectrometer. The converted interference signal was coupled by a single-mode fibre to a 4096-element array spectrometer, which allowed for high-resolution detection of a 1.3 um broad bandwidth covering 3.5-4.8~$\mu$m, at a line rate of 5 kHz. The system achieved a high axial (depth) resolution of  $\sim$ 7~$\mu$m, due to the large bandwidth of the laser, and a transverse spatial resolution of $\sim$ 30~$\mu$m,  limited by the lens magnification and a combination of spherical and chromatic aberrations. The sensitivity  6 dB roll-off was measured to be 1.39 mm with a maximum of $\sim$ 60 dB of sensitivity. For more information about the MIR OCT system see \cite{Israelsen2019,Israelsen2021}.\newline

The scan parameters for the characterisation were the following: 400 B-scans composed of 400

\begin{figure}[H]
\centering\includegraphics[width=15cm]{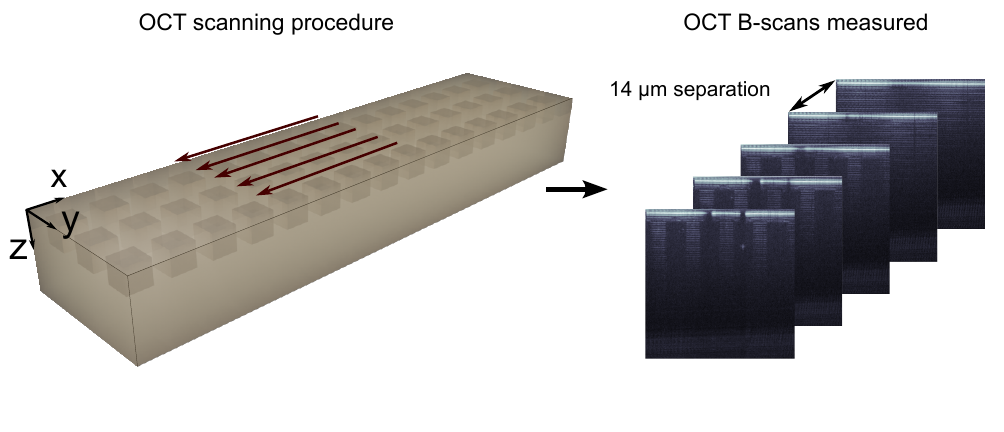}
\caption{Schematic of the OCT scanning procedure applied on the sample.} 
\label{scanning_procedure}
\end{figure}

A-scans were recorded, covering an area of 5.67$\times$5.67 mm$^2$, with each A-scan and B-scan being separated by $\sim$14~$\mu$m. Each B-scan took 200~$\mu$s to record. As the lens adds curvature to the cross-sectional image, a flattening post processing procedure was applied for correcting this. Finally, to increase image contrast, B-scan averaging was implemented. The averaging was based on five B-scans which covering $\sim$ 56~$\mu$m distance in the y-dimension, as illustrated in Fig.~\ref{scanning_procedure} .\newline

\section{Results}

\begin{figure}[H]
\centering\includegraphics[width=17cm]{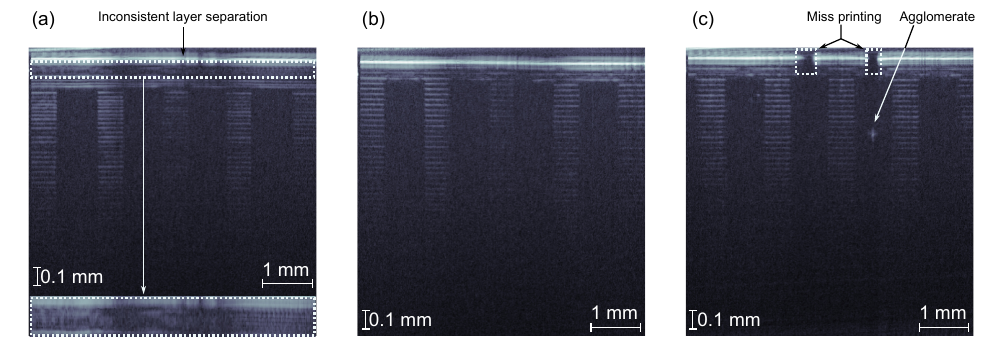}
\caption{Average of five B-scans of green body samples from (a) Group 1,   (b) Group 2, and (c) Group 3 (cf. Table~\ref{print_method}). }
\label{green}
\end{figure}

\begin{figure}[H]
\centering\includegraphics[width=17cm]{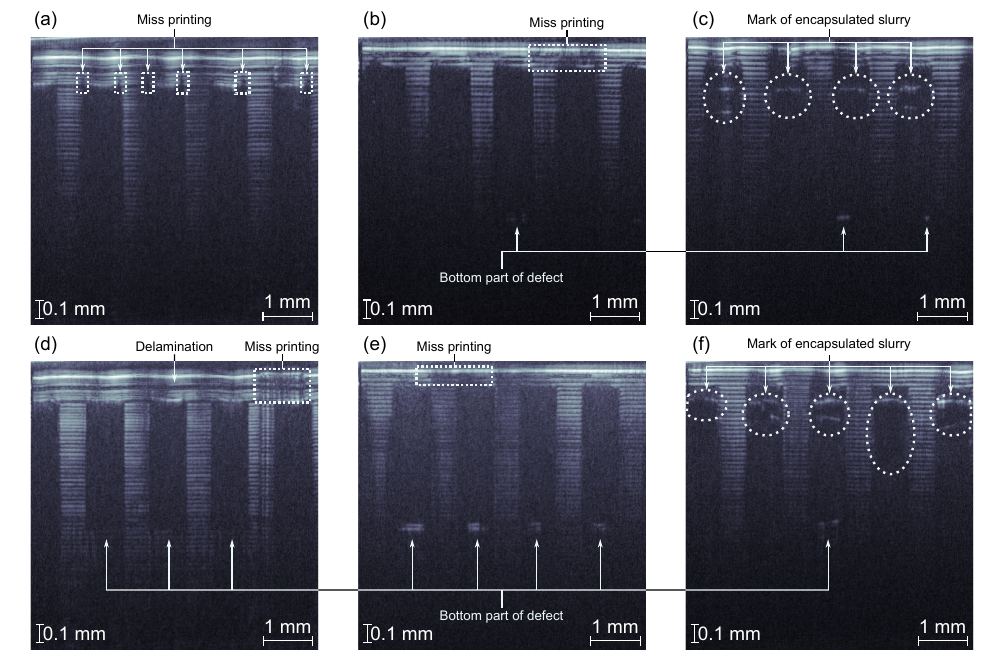}
\caption{Average of five B-scans of preconditioned samples from (a) group 1,   (b) group 2, and (c) group 3, and of debinded sample from (d) group 1,   (e) group 2, and (f) group 3  (cf. Table~\ref{print_method}). }
\label{precon+debinded}
\end{figure}

\begin{figure}[H]
\centering\includegraphics[width=17cm]{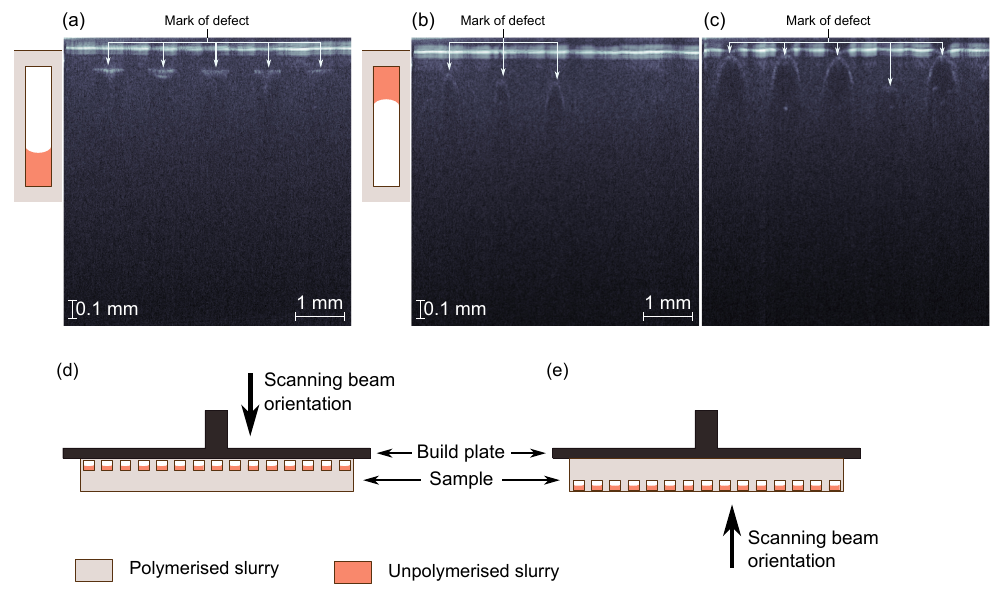}
\caption{Average of five B-scans of sintered samples from (a) group 1, (b) group 2, and (c) group 3 (cf. Table~\ref{print_method}). Under the B-scans schematics of the print orientation of the defects and how the slurry is positioned inside them is illustrated. (d) Defect on the top as group 1, (e) defect on the bottom as group 2 and 3. The thickness of the unpolymerised slurry is not representative.} 
\label{sintered}
\end{figure}

Figures~\ref{green}-\ref{sintered} show B-scans of samples from each processing step from each group. By comparing images from Groups 1 and 2 using the same beige slurry we can compare defects occurring in the top part of the sample (viewed from above) with defects occurring in the bottom part of the sample (viewed from below). By comparing images from Groups 2 and 3 we can compare defects occurring in samples made using two different beige and red slurries.

\begin{figure}[H]
\centering\includegraphics[width=15cm]{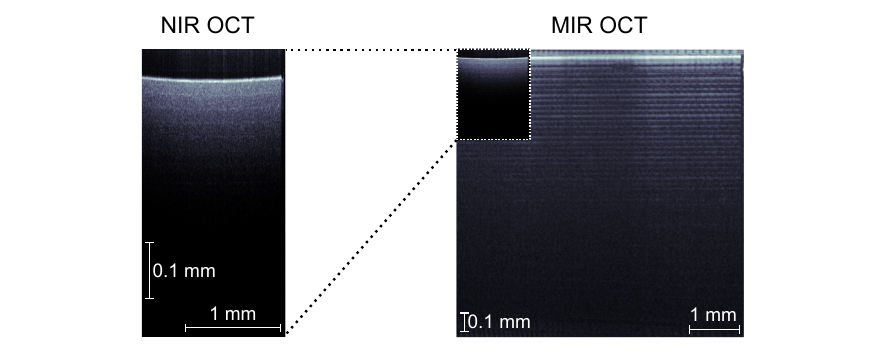}
\caption{B-scan comparison between NIR OCT (1.3~$\mu$m center wavelength) and MIR OCT (4~$\mu$m center wavelength). The dashed box in the MIR OCT B-scan presents the equally scaled NIR OCT B-scan.}
\label{compa_nir_mir}
\end{figure}

Each B-scan is composed of 400 A-scans, and the focus position was on the surface at the middle of the B-scans. The physical distance $L$ travelled by the light is linked to the so-called Optical Path Length ($OPL$) through the refractive index of the material it propagates in. It is the $OPL$ that an OCT system measures, so to get the actual physical distance we need to know the refractive index $n$. If the material is isotropic $OPL = n\times L$. We assumed that the ceramic material has an average refractive index of around n=1.67 \cite{malitson1972refractive}. With this assumption we could convert the observed OPL to physical length and found that we could observe structures until a depth of $\sim$ 0.7~mm, which corresponds to 29 print layers. This penetration was not possible to achieve with NIR OCT (1.3~$\mu$m center wavelength) which allows a depth penetration of only $\sim$ 0.1~mm, as shown in Fig~\ref{compa_nir_mir} \cite{Su2014,Israelsen2019,Zorin2022}.
\newline

Figure~\ref{green} shows the green sample from the three different groups, where images are averages over five B-scans to improve image contrast as explained in Section~\ref{mir_oct_section}. The green samples (cf. Section~\ref{alumina_section}) are the samples, which are cleaned from residual slurry around the part just after the print. In Fig.~\ref{green}(a), for a beige slurry and defects in the top part in the first number of layers of the printed sample, there seems to be an inconsistent layer separation in the top layers, $\sim$ 1-10 layers, before the defects were introduced. This type of inconsistency is not visible in the layers $\sim$ 90-102 being printed after the defects when they were introduced in the bottom part of the sample, as seen in Fig.~\ref{green}(b), which is also for the beige slurry. It is also noticeable that the  choice of slurry affects the quality of the final product. In  Fig.~\ref{green}(c), where the red slurry material is applied (Group 3), some areas right above the intentional defect were not printed due to a problem of incomplete polymerisation during the print. This type of defect is not observed for the beige slurry, as seen in the comparable Fig.~\ref{green}(b) which also had the defects in the final bottom part of the printing around layer 100. In addition, it was also possible to detect some authentic defects, such as agglomerates, as depicted in Fig.~\ref{green}(c). 
\newline

After cleaning the parts, the samples were subjected to three thermal post-processing steps. The first two were used to remove all the residual organic compounds, and the last one was used to densify the ceramic. Figures~\ref{precon+debinded}(a-c) and Fig.~\ref{precon+debinded}(d-f) show average B-scans from the three different groups of the preconditioned and debinded samples, respectively. 
\newline

As this process was used to remove the remaining organic part, the effect of the choice of slurry and the position of the defects, i.e., whether they occur during the printing of the first top layers or the final bottom layers, is more visible.  In Fig.~\ref{precon+debinded}, slurry seems to be encapsulated, (dashed ellipse inside Figs.~\ref{precon+debinded}(c,f) and Figs.~\ref{precon+debinded}(b,c)), and voids appear (dashed rectangles in Fig.~\ref{precon+debinded}(a) and  Fig.~\ref{precon+debinded}(b)). As the organic material encapsulated inside the defect started to solidify, it becomes possible to penetrate deeper into the sample and see the other deeper interface of some of the defects (cf. Figs.~\ref{precon+debinded}(b,c,d,e,f) ).
\newline

The B-scans of group 1 samples, shown in Figs.~\ref{precon+debinded}(a,d), reveal that the top surface has started  to collapse into the intentional defects when they appear in the early top layers of the printing. The (bottom) surface of both the preconditioned and debinded samples of group 3 also appears to have been influenced by the intentional defects, as is visible from the B-scans in Figs.~\ref{precon+debinded}(c,f).

Figures~\ref{sintered}(a-c) shows the sintered phase of each sample group. As the parts were sintered at 1650 °C for two hours, all layer structures disappeared, but marks of the intentional defects are still visible as indicated by arrows. \\

In Fig.~\ref{sintered}(a)  the entrance to the intentional defect is still visible in the OCT image as a straight interface and thus at least part of the defect is still present in the sintered sample. In Figs.~\ref{sintered}(b,c) the intentional defects are also still visible, but here the top entrance is curved as an ellipse. For Group 1 in Fig.~\ref{sintered}(a) the sample is imaged having the same orientation as it was printed, so the defects are still in the top part and gravity points down. In Groups 2-3 the defects were made in the bottom part of the sample, so during imaging we have rotated the samples to have the defects in the top part and therefore be best visible with OCT imaging as illustrated in Figs.~\ref{sintered}(d-e). This of course means that gravity is pointing up. The curved surfaces are therefore simply a reflection of gravity.\\

Indeed, due to the surface tension of slurry inside the defect, a curved interface appear in Fig.~\ref{sintered}(b,c) after the final process of sintering. In contrary, as seen in Fig.~\ref{sintered}(a), the interface of the defect is completely flat. The schematics at the left of Fig.~\ref{sintered}(a,b) show where the unpolymerised slurry was just after the print. Another interesting effect appear in Fig.~\ref{sintered}(c). The defect interface is closer to the surface than in the sample in  Fig.~\ref{sintered}(b). This difference can be explained by the difference in slurry recipe applied.

\section{Discussion and conclusion}

This study demonstrates the potential of MIR OCT as an NDI detection technique in-line characterisation of alumina printed samples with an array of intentional defects (air pockets) implemented into them. We used MIR OCT to study three groups of samples. The sub-surface imaging provided valuable information on the internal structure and it was affected by the defects and whether they occurred in the early top or final bottom part of the printed multi-layer structure. We identified the interior structures of the different samples to a volume depth level of about 0.7~mm, corresponding to about 30 print layers, and documented how air pockets in the green state have an impact on the final product. Even after all the thermal post-processing steps, the intentional defects still presented a significant impact on the final volume quality of the print. 
\newline

In the characterisation, we focused on the detection of intentional defects in three groups of different samples for the four phases of cleaning and thermal treatment: green, preconditioned, debinded, and sintered components. The cleaning and thermal post-processing followed the conventional LITHOZ tailored printing process. The difference between the three groups came from the choice of printing material (slurry) and on when in the printing process the defects appeared, - whether it was in the top early number of layers or in the bottom late number of layers. We were able to follow the impact of these difference (cf. Table~\ref{print_method}) during all the different phases of cleaning and thermal post-processing. The OCT images revealed distinct effects of the chosen print parameters, with the appearance of voids, collapsing surface, or layer separation, but also the presence of encapsulated unpolymerised slurry inside the defect. In addition, we also noticed unintended defects as a result of the print configuration, such as agglomerates inside the parts. \newline

The results of the MIR OCT scanner are encouraging for NDI characterisation given the detail level in the produced images. We measured an area of 5.67$\times$5.67~mm$^2$ in 1 min and 44 s. This time is slightly more than the minimum expected time of 1 min and 20 s due to the scanner movement. With an optimised scanning pattern, the measurement time could be reduced. The scanning area could be easily improved. To accommodate the total print area, a translation stage integration of the scanner head may be implemented. Scanning the whole depth dimension of the print can be addressed by either doing multiple scans at the same lateral positions, each scan after an appropriate number of printed layers, to ensure all layers are inspected. Indeed, most of the printed parts are larger than a cubic centimetre, and the actual penetration and scanning area, 76 $\times$ 43 $\times$ 170 mm$^3$ for the smallest and  250 $\times$ 250 $\times$ 290 mm$^3$ for the largest.  Additionally, at the cost of reducing the image acquisition speed, the spectrometer integration time can be increased to provide penetration beyond the reported depth of 0.7 mm. \newline

The MIR-OCT system used in this study employs a so-called upconversion spectrometer, making it ultra-fast with an A-scan rate of 5 kHz, while maintaining excellent penetration and depth resolution. One way to further improve the acquisition speed could be to utilise the so-called time-stretch technology, as was recently demonstrated by a group from Tokyo University, achieving an A-scan rate of 1 MHz \cite{Yagi2023}. However, both the sensitivity, the roll-off, and the depth resolution of that time stretch MIR OCT system were all below the performance of the up-conversion system presented here.
\\

In conclusion, this study highlights the impact of MIR OCT powered NDI in understanding volume AM material fabrication of a chosen alumina print and slurry recipe. The MIR OCT scanner enables us to follow all the different steps of fabrication and gives new understanding of existing defects and how they transform during the individual processing stages.\newline

\section*{Funding}

This project has received funding from Villum Fonden (2021 Villum Investigator project No. 00037822: Table-Top Synchrotrons).
\\
This project has received funding from Horizon Europe, the European Union’s Framework Programme for Research and Innovation, under Grant Agreement No. 101057404 (ZDZW), and Grant Agreement No. 101058054 (TURBO). 
Views and opinions expressed are however those of the authors only and do not necessarily reflect those of the European Union. The European Union cannot be held responsible for them.
\\

\section*{Data and Code availability}
The data presented in this study are available on request from the
corresponding author. The data are not publicly available due to significant storage requirements.

\section*{Acknowledgements}
We would like to acknowledge María Rocío Del Amor, Fernando García Torres, Natalia Lourdes Perez Garcia de la Puent, Adrian Colomer Granero and Prof. Valery Naranjo of the CVB lab, la Universitat Politècnica de València for rewarding discussions on localization and annotation of material defects observed in OCT images.

\section*{Competing interests}
The authors declare that they have no known competing financial interests or personal relationships that could have appeared to influence the work reported in this paper.

\bibliography{lapre}

\begin{thebibliography}{10}

\bibitem{Wang2017}
X.~Wang, M.~Jiang, Z.~Zhou, J.~Gou, and D.~Hui, ``{3D printing of polymer matrix composites: A review and prospective},'' {\em Composites Part B: Engineering}, vol.~110, pp.~442--458, Feb. 2017.

\bibitem{Shahrubudin2019}
N.~Shahrubudin, T.~Lee, and R.~Ramlan, ``{An Overview on 3D Printing Technology: Technological, Materials, and Applications},'' {\em Procedia Manufacturing}, vol.~35, pp.~1286--1296, 2019.

\bibitem{Benamira2023}
M.~Benamira, N.~Benhassine, A.~Ayad, and A.~Dekhane, ``Investigation of printing parameters effects on mechanical and failure properties of 3d printed pla,'' {\em Engineering Failure Analysis}, vol.~148, p.~107218, June 2023.

\bibitem{Xiao2021}
J.~Xiao, G.~Ji, Y.~Zhang, G.~Ma, V.~Mechtcherine, J.~Pan, L.~Wang, T.~Ding, Z.~Duan, and S.~Du, ``{Large-scale 3D printing concrete technology: Current status and future opportunities},'' {\em Cement and Concrete Composites}, vol.~122, p.~104115, Sept. 2021.

\bibitem{Quan2020}
H.~Quan, T.~Zhang, H.~Xu, S.~Luo, J.~Nie, and X.~Zhu, ``{Photo-curing 3D printing technique and its challenges},'' {\em Bioactive Materials}, vol.~5, pp.~110--115, Mar. 2020.

\bibitem{Su2008}
B.~Su, S.~Dhara, and L.~Wang, ``Green ceramic machining: A top-down approach for the rapid fabrication of complex-shaped ceramics,'' {\em Journal of the European Ceramic Society}, vol.~28, pp.~2109--2115, Jan. 2008.

\bibitem{Hassanin2021}
H.~Hassanin, K.~Essa, A.~Elshaer, M.~Imbaby, H.~H. El-Mongy, and T.~A. El-Sayed, ``Micro-fabrication of ceramics: Additive manufacturing and conventional technologies,'' {\em Journal of Advanced Ceramics}, vol.~10, pp.~1--27, Jan. 2021.

\bibitem{Steckenrider2013}
J.~S. Steckenrider, J.~Torcedo, and J.~Kutsch, ``Multimodal characterization of transparent dome blanks,'' in {\em Window and Dome Technologies and Materials XIII} (R.~W. Tustison and B.~J. Zelinski, eds.), SPIE, June 2013.

\bibitem{Duan2019}
Y.~Duan, H.~Zhang, S.~Sfarra, N.~P. Avdelidis, T.~H. Loutas, G.~Sotiriadis, V.~Kostopoulos, H.~Fernandes, F.~I. Petrescu, C.~Ibarra-Castanedo, and X.~P. Maldague, ``{On the Use of Infrared Thermography and Acousto—Ultrasonics NDT Techniques for Ceramic-Coated Sandwich Structures},'' {\em Energies}, vol.~12, p.~2537, July 2019.

\bibitem{Zhao2021}
Z.~Zhao, ``Review of non-destructive testing methods for defect detection of ceramics,'' {\em Ceramics International}, vol.~47, pp.~4389--4397, feb 2021.

\bibitem{Drexler2001}
W.~Drexler, U.~Morgner, R.~K. Ghanta, F.~X. Kärtner, J.~S. Schuman, and J.~G. Fujimoto, ``Ultrahigh-resolution ophthalmic optical coherence tomography,'' {\em Nature Medicine}, vol.~7, pp.~502--507, Apr. 2001.

\bibitem{Israelsen2018}
N.~M. Israelsen, M.~Maria, M.~Mogensen, S.~Bojesen, M.~Jensen, M.~Haedersdal, A.~Podoleanu, and O.~Bang, ``The value of ultrahigh resolution {OCT} in dermatology - delineating the dermo-epidermal junction, capillaries in the dermal papillae and vellus hairs,'' {\em Biomedical Optics Express}, vol.~9, p.~2240, apr 2018.

\bibitem{Fercher1988}
A.~F. Fercher, K.~Mengedoht, and W.~Werner, ``Eye-length measurement by interferometry with partially coherent light,'' {\em Optics Letters}, vol.~13, p.~186, mar 1988.

\bibitem{Fujimoto2000}
J.~G. Fujimoto, C.~Pitris, S.~A. Boppart, and M.~E. Brezinski, ``{Optical Coherence Tomography: An Emerging Technology for Biomedical Imaging and Optical Biopsy},'' {\em Neoplasia}, vol.~2, pp.~9--25, jan 2000.

\bibitem{Boehringer2009}
H.~J. Böhringer, E.~Lankenau, F.~Stellmacher, E.~Reusche, G.~Hüttmann, and A.~Giese, ``Imaging of human brain tumor tissue by near-infrared laser coherence tomography,'' {\em Acta Neurochirurgica}, vol.~151, pp.~507--517, apr 2009.

\bibitem{Walther2011}
J.~Walther, M.~Gaertner, P.~Cimalla, A.~Burkhardt, L.~Kirsten, S.~Meissner, and E.~Koch, ``Optical coherence tomography in biomedical research,'' {\em Analytical and Bioanalytical Chemistry}, vol.~400, pp.~2721--2743, may 2011.

\bibitem{Petersen2021}
C.~R. Petersen, N.~Rajagopalan, C.~Markos, N.~M. Israelsen, P.~J. Rodrigo, G.~Woyessa, P.~Tidemand-Lichtenberg, C.~Pedersen, C.~E. Weinell, S.~Kiil, and O.~Bang, ``{Non-Destructive Subsurface Inspection of Marine and Protective Coatings Using Near- and Mid-Infrared Optical Coherence Tomography},'' {\em Coatings}, vol.~11, p.~877, jul 2021.

\bibitem{Hansen2022}
R.~E. Hansen, T.~B{\ae}k, S.~L. Lange, N.~M. Israelsen, M.~Mäntylä, O.~Bang, and C.~R. Petersen, ``{Non-Contact Paper Thickness and Quality Monitoring Based on Mid-Infrared Optical Coherence Tomography and {THz} Time Domain Spectroscopy},'' {\em Sensors}, vol.~22, p.~1549, feb 2022.

\bibitem{Petersen2023}
C.~R. Petersen, S.~Fæster, J.~I. Bech, K.~M. Jespersen, N.~M. Israelsen, and O.~Bang, ``Non‐destructive and contactless defect detection inside leading edge coatings for wind turbine blades using mid‐infrared optical coherence tomography,'' {\em Wind Energy}, vol.~26, pp.~458--468, Mar. 2023.

\bibitem{Zorin2022}
I.~Zorin, D.~Brouczek, S.~Geier, S.~Nohut, J.~Eichelseder, G.~Huss, M.~Schwentenwein, and B.~Heise, ``Mid-infrared optical coherence tomography as a method for inspection and quality assurance in ceramics additive manufacturing,'' {\em Open Ceramics}, vol.~12, p.~100311, Dec. 2022.

\bibitem{ZDZW_project}
{ZDZW project}, ``Non-destructive inspection services for digitally enhanced zero waste manufacturing \url{{https://www.zdzw-project.eu/}},'' 2024.

\bibitem{Conti2020}
L.~Conti, D.~Bienenstein, M.~Borlaf, and T.~Graule, ``{Effects of the Layer Height and Exposure Energy on the Lateral Resolution of Zirconia Parts Printed by Lithography-Based Additive Manufacturing},'' {\em Materials}, vol.~13, p.~1317, Mar. 2020.

\bibitem{LithozGmbH2022}
L.~GmbH, ``{Highest Resolution from a 3D Printer},'' {\em Interceram - International Ceramic Review}, vol.~71, pp.~16--17, Dec. 2022.

\bibitem{Ferkel1999}
H.~Ferkel and R.~Hellmig, ``Effect of nanopowder deagglomeration on the densities of nanocrystalline ceramic green bodies and their sintering behaviour,'' {\em Nanostructured Materials}, vol.~11, pp.~617--622, Aug. 1999.

\bibitem{woyessa2021power}
G.~Woyessa, K.~Kwarkye, M.~K. Dasa, C.~R. Petersen, R.~Sidharthan, S.~Chen, S.~Yoo, and O.~Bang, ``Power stable 1.5--10.5 $\mu$m cascaded mid-infrared supercontinuum laser without thulium amplifier,'' {\em Optics Letters}, vol.~46, no.~5, pp.~1129--1132, 2021.

\bibitem{Wojtkowski2004}
M.~Wojtkowski, V.~J. Srinivasan, T.~H. Ko, J.~G. Fujimoto, A.~Kowalczyk, and J.~S. Duker, ``{Ultrahigh-resolution, high-speed, Fourier domain optical coherence tomography and methods for dispersion compensation},'' {\em Optics Express}, vol.~12, p.~2404, May 2004.

\bibitem{Rottwitt2014}
K.~Rottwitt and P.~Tidemand-Lichtenberg, {\em {Nonlinear Optics Principles and Applications}}.
\newblock Taylor {\&} Francis Group, 2014.

\bibitem{Israelsen2019}
N.~M. Israelsen, C.~R. Petersen, A.~Barh, D.~Jain, M.~Jensen, G.~Hannesschläger, P.~Tidemand-Lichtenberg, C.~Pedersen, A.~Podoleanu, and O.~Bang, ``Real-time high-resolution mid-infrared optical coherence tomography,'' {\em {Light: Science {\&} Applications}}, vol.~8, jan 2019.

\bibitem{Israelsen2021}
N.~M. Israelsen, P.~J. Rodrigo, C.~R. Petersen, G.~Woyessa, R.~E. Hansen, P.~Tidemand-Lichtenberg, C.~Pedersen, and O.~Bang, ``{High-resolution mid-infrared optical coherence tomography with kHz line rate},'' {\em Optics Letters}, vol.~46, p.~4558, Sept. 2021.

\bibitem{malitson1972refractive}
I.~H. Malitson and M.~J. Dodge, ``Refractive index and birefringence of synthetic sapphire,'' {\em J. Opt. Soc. Am}, vol.~62, no.~11, p.~1405, 1972.

\bibitem{Su2014}
R.~Su, M.~Kirillin, E.~W. Chang, E.~Sergeeva, S.~H. Yun, and L.~Mattsson, ``Perspectives of mid-infrared optical coherence tomography for inspection and micrometrology of industrial ceramics,'' {\em Optics Express}, vol.~22, p.~15804, June 2014.

\bibitem{Yagi2023}
S.~Yagi, T.~Nakamura, K.~Hashimoto, S.~Kawano, and T.~Ideguchi, ``{Mid-infrared optical coherence tomography with MHz axial line rate for real-time non-destructive testing},'' {\em arXiv}, 2023.

\end{thebibliography}
\bibliographystyle{ieeetr}

\end{document}